# $M^2$ANET: Mobile Malaria Attention Network for efficient classification of plasmodium parasites in blood cells

Salam Ahmed Ali[1], Peshraw Salam Abdulqadir[1], Shan Ali Abdullah[2] & Haruna Yunusa[3]

[1] University of Garmian, [2] Sulaimani Polytechnic University, [3] Beihang University

**Abstract.** Malaria is a life-threatening infectious disease caused by Plasmodium parasites, which poses a significant public health challenge worldwide, particularly in tropical and subtropical regions. Timely and accurate detection of malaria parasites in blood cells is crucial for effective treatment and control of the disease. In recent years, deep learning techniques have demonstrated remarkable success in medical image analysis tasks, offering promising avenues for improving diagnostic accuracy, with limited studies on hybrid mobile models due to the complexity of combining two distinct model and the significant memory demand of self-attention mechanism especially for edge devices. In this study, we explore the potential of designing a hybrid mobile model for efficient classification of plasmodium parasites in blood cell images. Therefore, we present $M^2$ANET (Mobile Malaria Attention Network). The model integrates MBConv3 (MobileNetV3 blocks) for efficient capturing of local feature extractions within blood cell images and a modified global-MHSA (multi-head self-attention) mechanism in the latter stages of the network for capturing global context. Through extensive experimentation on benchmark, we demonstrate that $M^2$ANET outperforms some state-of-the-art lightweight and mobile networks in terms of both accuracy and efficiency. Moreover, we discuss the potential implications of $M^2$ANET in advancing malaria diagnosis and treatment, highlighting its suitability for deployment in resource-constrained healthcare settings. The development of $M^2$ANET represents a significant advancement in the pursuit of efficient and accurate malaria detection, with broader implications for medical image analysis and global healthcare initiatives.

**Keywords:** Attention mechanism, Computer Vision, Malaria Detection, Medical Image Analysis, Mobile Hybrid Model

## 1 Introduction

The detection and diagnosis of malaria, a life-threatening infectious disease caused by Plasmodium parasites transmitted through the bites of infected mosquitoes, remain critical challenges in global healthcare [1]. With millions of cases reported annually, particularly in tropical and subtropical regions, the timely and accurate identification of malaria parasites in blood cells is paramount for effective treatment and control [2]. However, the current method of malaria diagnosis relies heavily on manual microscopic examination of the appearance, number, and shape of red blood cells. This approach necessitates the involvement of skilled medical experts, rendering the process both time-consuming and costly [3, 4, 5, 6, 7]. Moreover, the subjective nature of visual examination means that diagnostic results may occasionally be inaccurate due to human errors, highlighting the pressing need for more reliable and efficient automated visual diagnostic method [8].

In recent years, computer vision techniques have demonstrated remarkable success in various medical image analysis tasks, including disease detection and diagnosis [9]. However, these methods are primarily deployed as computer-aided diagnostic (CAD) systems to provide rapid assistance and enhance the accuracy of disease diagnosis [10]. Previous research has focused on classifying and detecting malaria in cells using conventional methods such as Convolutional Neural Network (CNN) architectures like LeNet [11], VGG [12], AlexNet [13], Inception [14], ResNet [15], and EfficientNet [16], as well as other non-CNN methods like Support Vector Machine (SVM) [17] and XG-Boost [18, 19]. While these approaches have shown good performance in detecting and diagnosing malaria diseases, the adoption of custom methods tailored to specific disease detection tasks holds the potential for even greater precision, efficiency, and reliability in diagnosis, thereby advancing the field of medical imaging and improving patient outcomes. Additionally, limited research has explored the utilization of self-attention mechanisms to enhance malaria disease detection.

The self-attention mechanism was initially developed for (NLP) natural language processing but later adapted for computer vision models to enhance capturing long-range dependencies and relationships within the input data [38]. This mechanism allows the model to dynamically focus on relevant regions of the image while considering the interactions between different parts of the input. It helps the model to understand the global context of an image, which is particularly beneficial when identifying complex patterns that require holistic analysis rather than just local features [37]. In medical image analysis, self-attention is especially important because it allows for more accurate identification of subtle and dispersed features within medical images, such as the irregular shapes and sizes of Plasmodium parasites in blood smears. By considering the entire image context, self-attention can enhance the detection of anomalies that might be missed by models relying solely on local information.

However, the application of self-attention in vision models, particularly in resource-constrained environments like mobile and edge devices, faces significant challenges. Self-attention mechanisms tend to be computationally intensive and memory-demanding, which limits their practicality in these settings [39]. Hybrid models, which combine CNNs with self-attention mechanisms, offer a promising solution to these limitations. By leveraging the efficient local feature extraction capabilities of CNNs and integrating them with the global context capturing power of self-attention, hybrid models can achieve a balance between computational efficiency and performance [37]. This approach ensures that the model remains lightweight and suitable for deployment on mobile and edge devices while maintaining high accuracy and robustness in medical image analysis. Figure 1 shows the performance of $M^2$ANET in classifying Plasmodium parasites in blood cell images compared to some recent methods in terms of accuracy and trainable parameters.

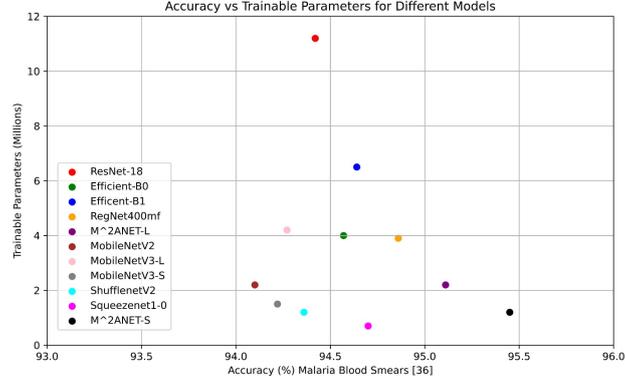

**Figure 1** $M^2$ANET comparison with recent methods

In this study, we design a novel model for detecting malaria in blood cells that classifies parasitized and unparasitized cells. The model is a hybrid mobile network using MBConv3 [20] in the first two stages for efficient features extraction and modified 2D MHSA (multi-head self-attention) in the latter stages for improved global context captures. We fused both layers using a pair-wise addition. The network is dubbed as $M^2$ANET (Mobile Malaria Attention Network). $M^2$ANET integrates attention mechanisms to dynamically highlight informative regions within cell images while maintaining computational efficiency. This integration allows $M^2$ANET to surpass conventional mobile models like MobileNets [20], ShuffleNet [21], SqueezeNet [22], etc. while remaining computationally efficient, making it suitable for deployment in resource constrained mobile and edge devices such as healthcare facilities in malaria endemic regions. We believe this study represents a significant step forward in the search for efficient, accurate and low-resource models for malaria disease detection, aiming to accurately diagnose this deadly disease.

The contributions of this paper are summarized as follows:
- We propose $M^2$ANET, an efficient mobile-based hybrid model for detecting malaria disease using red blood cells images.
- The model is designed for mobile and edge devices which is computationally efficient in real time.
- The model can serve as a baseline hybrid for identifying plasmodium parasites in blood cells images, where developers and researcher can continue to explore the synergy of employing two distinct models efficiently.

## 2    Related Work

**Deep Learning Methods.** Previous studies have been conducted to detect malaria in blood cells with promising results using deep learning methods including Deep Belief Network (DBN) and CNN. The research by Bibin et al. (2017) [23] proposed a novel method utilizing DBN for detecting malaria parasites in peripheral blood smear images, achieving high accuracy through pre-training with contrastive divergence. While, Sivaramakrishnan et al. (2017) [24] proposed a custom CNN model for malaria cell classification, achieving a high accuracy of 98.61% by visualizing features and activations within the model. Then again, Sivaramakrishnan et al. (2018) [25] evaluated pre-trained CNNs for malaria parasite detection in blood smear images, demonstrating promising results for feature extraction and classification. Yang et al. (2019) [26] developed a deep learning method for automated malaria parasite detection in thick blood smear images using smartphones, achieving high accuracy and correlation with ground truth. Vijayalakshmi et al. (2020) [27] introduced a deep neural network model for identifying infected falciparum malaria parasites using transfer learning with VGG-SVM, outperforming existing CNN models in accuracy and performance indicators. Loddo et al. (2022) [28] investigated deep learning architectures for malaria diagnosis, comparing conventional CNN models and evaluating their performance on different datasets, highlighting the need for further research to improve robustness. Madhu et al. (2021) [29] developed a Deep Siamese Capsule Network (D-SCN) model for automatic diagnosis of malaria parasites in thin blood smears, achieving high detection and classification accuracy. Siłka et al. (2023) [30] proposed an AI-based object detection system for malaria diagnosis, achieving high accuracy comparable to human microscopists, which could aid diagnosis in resource-limited regions. Abdurahman et al. (2021) [31] investigated modified YOLOV3 and YOLOV4 models for malaria parasite detection in thick blood smear images, achieving state-of-the-art accuracy and outperforming other detection methods. Zhong et al. (2023) [32] developed an efficient malaria detection system using CNN adapted for mobile devices and microscopes, achieving high accuracy with diverse image dataset of various regions.

**Machine Learning Methods.** Aris et al. (2020) [33] propose a fast k-means clustering algorithm for malaria detection in thick blood smear images, evaluating 5 color models and 15 color components. Their study concludes that segmentation through the R component of RGB achieves the highest accuracy at 99.81%. Jahan & Alam (2023) [34] introduce a hybrid machine learning algorithm to classify malaria-infected erythrocytes, combining supervised algorithms such as stochastic gradient descent, logistic regression, decision trees, and XGBoost. Python-based approach achieves 95.64% and 96.22% accuracy in two configurations, aiding medical practitioners in malaria diagnosis. Murmu & Kumar (2024) [35] present DLRFNet, combines deep CNN with Random Forest to detect plasmodium malaria parasite. Their method address data scarcity and imbalance, improving on existing models and shows it effectiveness in parasite detection, leverages domain-specific expertise and integrating modifications for enhanced visualization and precise boundary detection.

**Discussion.** The review shows a notable lack of exploration into the use of self-attention mechanisms and hybrid mobile model designs for detecting malaria in blood smears. Furthermore, many existing models exhibit high computational costs, rendering them unsuitable for deployment on mobile, edge devices or low-power embedded systems. Our work fills this gap by effectively combining CNNs with self-attention mechanisms to detect malaria-infected red blood cells. Importantly, our design is well-suited for use on mobile and edge devices, offering a promising solution to the limitations of existing methods.

## 3     Method

### 3.1     $M^2$ANET Overview

$M^2$ANET is a novel hybrid mobile deep learning model designed to enhance the classification accuracy of blood cell images, particularly in classifying between plasmodium parasitized and non-parasitized cells. It achieves this through the combination of spatial feature extraction capabilities from MBConv3, based on the MobileNetV3 architecture, and a modified 2D global MHSA. The modification introduces a grouped point-wise convolution to the query, key and value projections in the MHSA, effectively reducing computational and memory complexity. This integration facilitates efficient processing in resource-constrained environments, such as mobile and edge devices, while precisely capturing local and global context within blood cell images, thereby improving the classification accuracy of plasmodium parasitized and non-parasitized cells, promising enhanced accuracy and efficiency in malaria diagnosis and treatment. Figure 2 shows architectural design of the model.

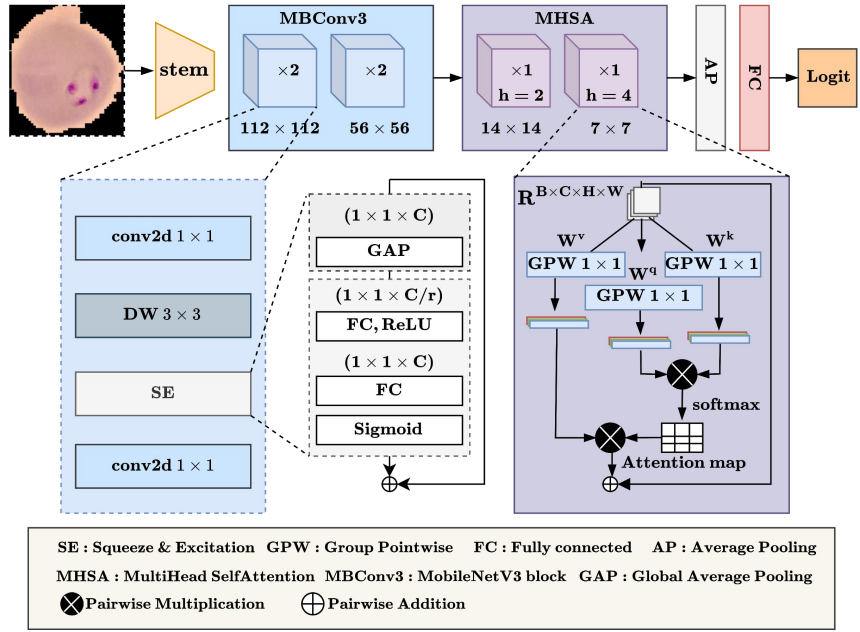

**Figure 2** $M^2$ANET architecture

### 3.2   $M^2$ANET Architecture

**Input.** $M^2$ANET processes 2D RGB images of size 112 × 112 pixels, with each image having to encode the color information, enabling the model to perceive visual features and patterns. The consistent small image size ensures uniform processing, and within the network and facilitates compatibility across various low-resourced edge devices. Standardizing the input dimensions allows $M^2$ANET to effectively analyze and extract relevant information, enabling efficient robust plasmodium parasite cell disease detection.

**Stem.** The stem block in $M^2$ANET serves as a basic feature extractor, aiming to extract fine-grained information from the input image before reducing its spatial dimensions. We preferred this approach over directly reducing the spatial dimension for computational efficiency, as the latter can lead to information loss. The stem block consists of a single layer of 3 × 3 convolution with batch normalization and ReLU activation, all with a stride of 1. Let $X$ denote the input Plasmodium cell image, where $X \in \mathbb{R}^{H \times W \times C}$.

**Efficient Local Details.** The section of this network aims to extract fine-grained local-features that is low on computational demand which will efficiently run on edge-devices and mobile devices. Therefore, we utilized the MBConv3 blocks (MobileNetV3) to serve this purpose since they are computationally efficient and built to work in mobile and edge devices. These blocks adopt a bottleneck design with blocks, a pairwise conv 1 × 1 for projection, then a 3 × 3 depth-wise separable convolution, then a squeeze & excitation (SE) layer for channel-wise calibration, and lastly a 1 × 1 conv for dimensionality. For our design we arranged these blocks as [2, 2] for local-details extraction. We represent the operations within the MBConv3 block from equation 1- 4.

Pairwise convolution (1 × 1 convolution)
$$\text{Proj}(x) = \text{Conv}_{1 \times 1}(x, W_{\text{proj}}) + b_{\text{proj}} \tag{1}$$

Depth-wise separable convolution
$$\text{DWConv}(x) = \text{Conv}_{3 \times 3}^{\text{depth-wise}}(x, W_{\text{DW}}) + b_{\text{DW}} \tag{2}$$

Squeeze and Excitation (SE) Layer
$$\text{SE}(x) = \sigma(\text{avgpool}\left(\text{ReLU}(\text{Conv}_{1 \times 1}(x, W_{\text{squeeze}}) + b_{\text{squeeze}})\right) \odot x \tag{3}$$

Dimensionality reduction

$$\text{DimRed}(x) = \text{Conv}_{1\times1}(x, W_{\text{dimred}}) + b_{\text{dimred}} \tag{4}$$

**Lightweight Global Details.** This section discusses the integration of 2D *global* MHSA into $M^2$ANET, emphasizing the need to reduce computational complexity for mobile devices. In the conventional approach, pointwise convolutions are applied to the query, key, and value projections within MHSA which is effective but can lead to significant computational overhead, particularly when handling high-dimensional inputs. To address this, we introduce the use of grouped pointwise convolutions, which maintain performance while substantially improving computational efficiency.

Grouped pointwise convolutions involve applying convolutions independently to groups of channels within the input tensor. This approach reduces the number of parameters and the overall computational complexity. In the standard configuration, each input channel interacts with every output channel, resulting in a computational burden. Specifically, for input tensor $X \in \mathbb{R}^{n \times c \times h \times w}$ with c channels, the pointwise convolution is applied as in equation 5.

$$Y = X * W + b \tag{5}$$

where, $W \in \mathbb{R}^{c \times c \times 1 \times 1}$ and $b \in \mathbb{R}^c$

To enhance efficiency, we modify the convolutions to be grouped. Each group processes a subset of the input channels independently. Given the same input tensor, the grouped convolution is applied with $g$ groups (where $g = c$), such that each group contains one channel as in equation 6.

$$Y = X * W + b \tag{6}$$

Where, $W_g \in \mathbb{R}^{(c/g) \times (c/g) \times 1 \times 1}$ and $b_g \in \mathbb{R}^{c/g}$

The choice of grouped pointwise convolutions is due to their ability to decouple the interactions between different channels. By processing each channel independently, the computational complexity is reduced from $O(C^2)$ to $O(C)$, where $C$ is the number of channels. This reduction in complexity translates to fewer parameters and operations, thereby improving computational efficiency while maintaining the expressiveness of the model. Thus, enhances the model's suitability for deployment on mobile devices. Despite the streamlined computations, the proposed method ensures that the performance of the attention mechanism remains robust. This balance between efficiency and performance is crucial for mobile classifiers that need to operate under resource constraints environment without compromising accuracy.

**Non-isotropic architecture.** Since $M^2$ANET is a hybrid model that integrates an attention mechanism which mostly functions in an isotropic architecture (i.e., maintaining the same feature spatial resolution throughout the whole depths with no down-sampling). However, this choice of design is computationally costly because it treats all input dimensions equally, regardless of their orientation or position. This means that isotropic architecture requires a larger number of parameters to learn the same level of complexity as non-isotropic architecture. Therefore, we adopted the pyramid-like structure of traditional CNNs. $M^2$ANET down-samples feature maps by applying a stride of 2 after each stage of the network, thus down-sampling the feature maps and significantly reducing the model's computational complexity, making it suitable for low-resource environments

**Interaction between Local and Global Features.** $M^2$ANET achieves a seamless fusion of local and global features within its architecture. This fusion involves combining the outputs from the MBConv3 blocks, representing local features ($L_{local}$), with the inputs to the 2D global MHSA, representing global features ($G_{global}$), equation 7.

$$F_{fused} = L_{local} \oplus G_{global} \tag{7}$$

Here, $\oplus$ denotes the fusion operation. This integration mechanism ensures that both fine-grained details and broader contextual information are effectively combined to achieve a comprehensive understanding of the input image.

## 4   Experimental Results and Comparison

This section presents the experimental results and comparisons with state-of-the-art mobile networks such as MobileNetV2 [20], MobileNetV3-L [20], MobileNetV3-S [20], ShufflenetV2 [21], Squeezenet1-0 [22], as well as lightweight models such as ResNet-18 [40], Efficient-B0 [41], Efficent-B1[41] and RegNet400mf [42]. The models were evaluated using malaria-infected thin blood smear images to classify parasitized and non-parasitized cells.

**Datasets**. The Malaria dataset comprises of 27,558 cell images, ranging from 150 to 150 pixels, evenly divided between 13,779 parasitized cell images and 13,779 uninfected cells images. These images are derived from thin blood smear slides of segmented cells [36].

**Experimental Configuration.** The experiments were performed on a Linux-based machine equipped with an Intel Core i7 8700k processor, two NVIDIA Titan XP 12GB GPUs, and 32GB of RAM. The training procedure included 90 epochs with a batch size of 64. The AdamW optimizer was used with an initial learning rate of 0.0001 and a weight decay rate of 0.05.

### 4.1   Grad-CAM Visualization

To visualize and evaluate $M^2$ANET ability to detect the Plasmodium parasite in thin blood smear images, we used the Grad-CAM method. This method allows us to observe the model's ability to focus on regions of infection, demonstrating good interpretability, which is crucial for understanding how the model makes decisions. We also compared the results with other state-of-the-art models; see figure 3.

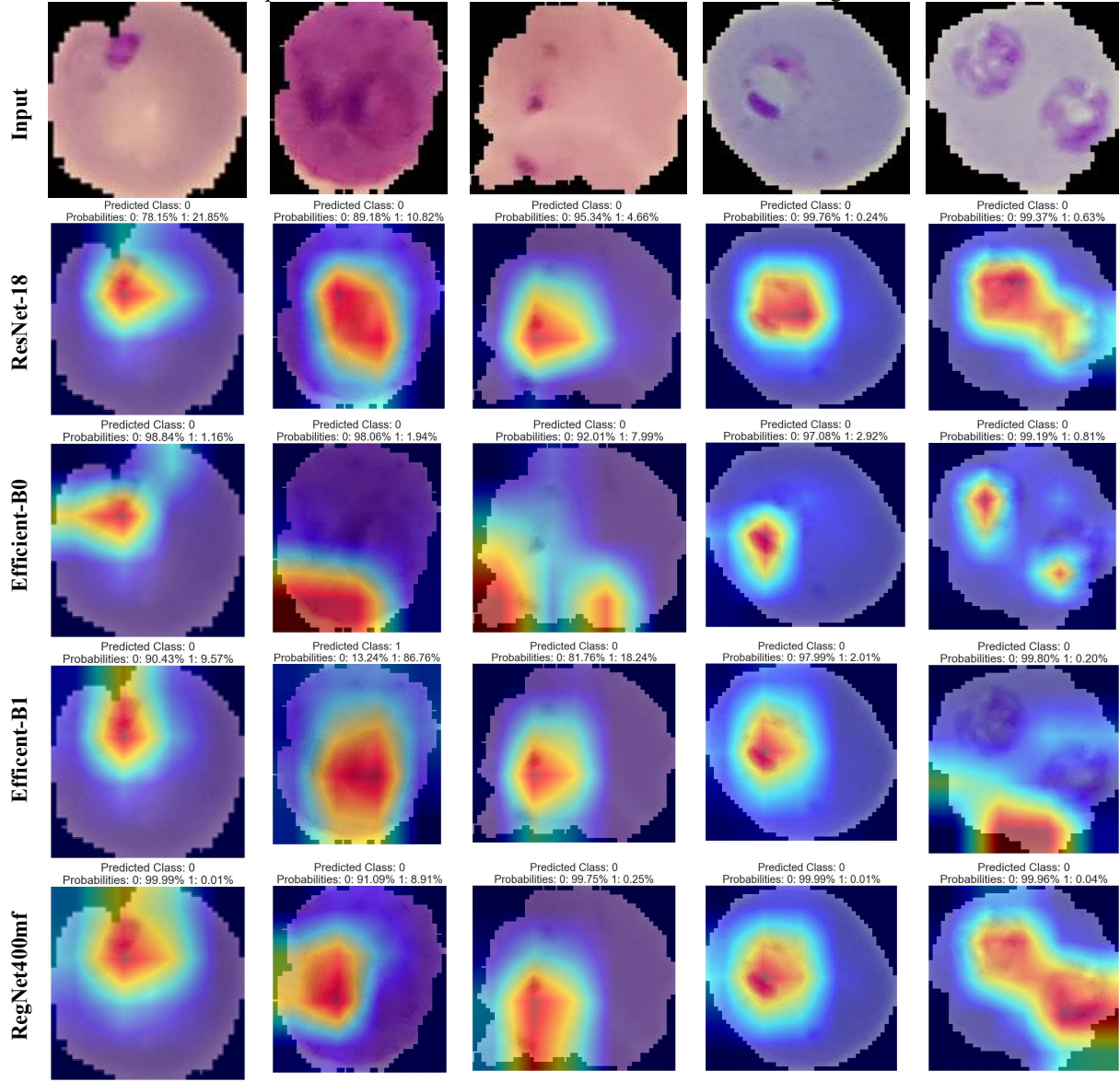

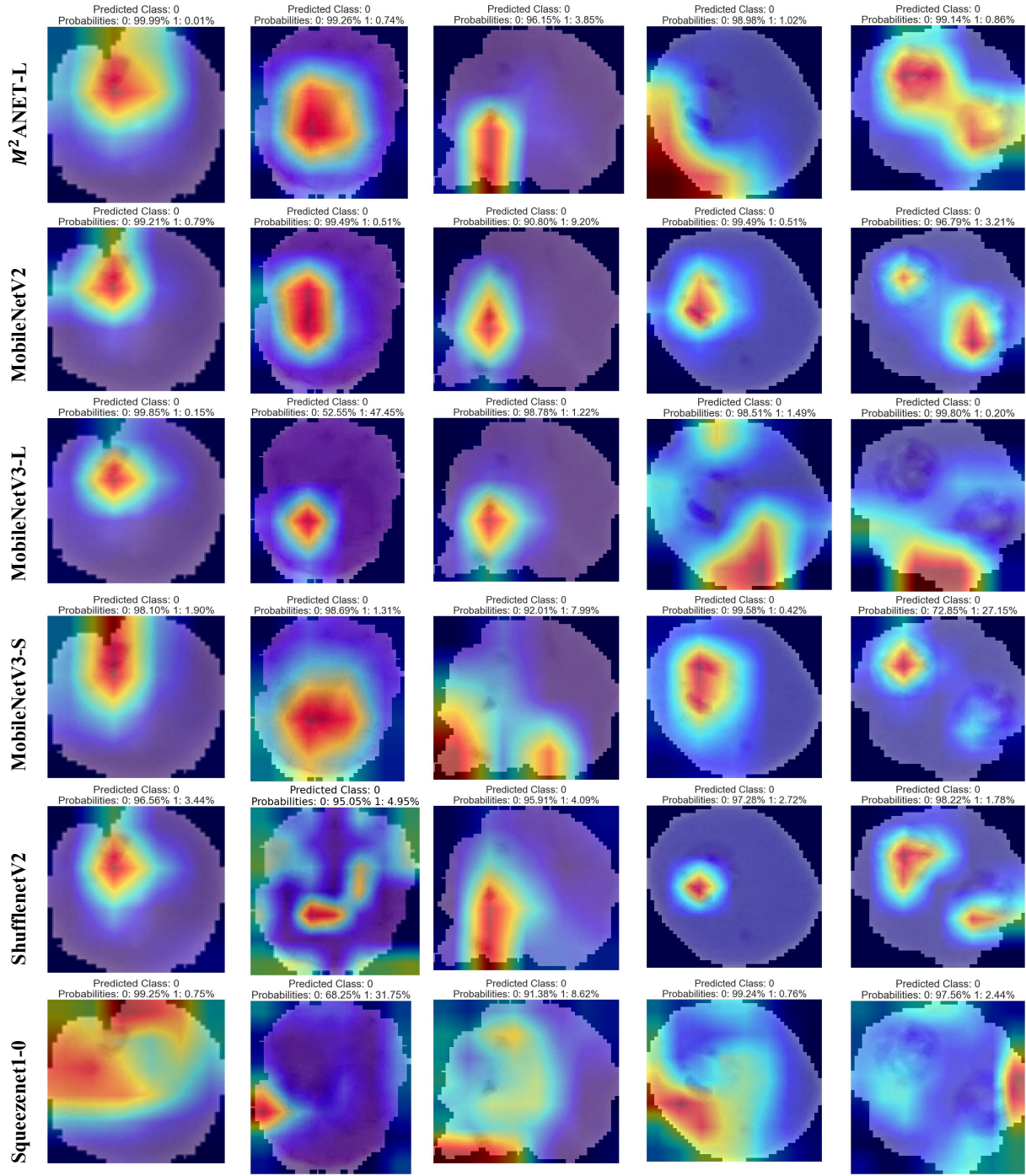

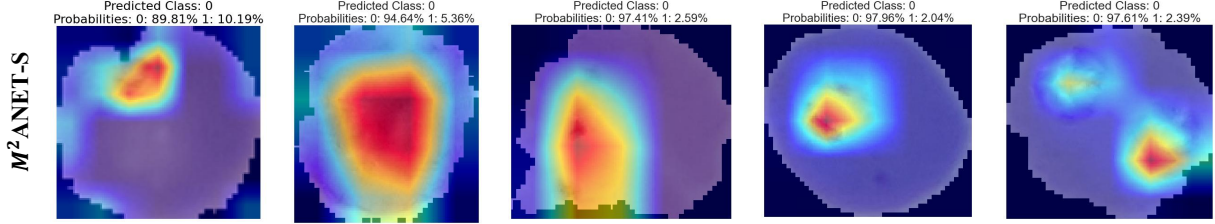

**Figure 3** Comparing GRAD-CAM visualization with several methods

**ROC & Precision-Recall Curve.** $M^2$ANET achieves an ROC value of 0.95 in figure 4, the highest among compared methods. This indicates its effectiveness in distinguishing infected cells from uninfected ones by maintaining a high true positive rate while minimizing false positives. Similarly, achieving a precision-recall curve score of 0.96 in figure 5 provides valuable insight into the trade-off between precision and recall, further highlighting the model's performance in identifying infected cells.

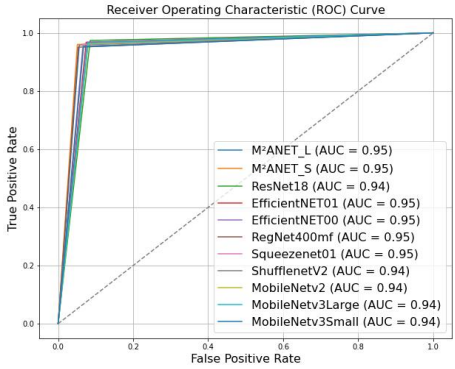
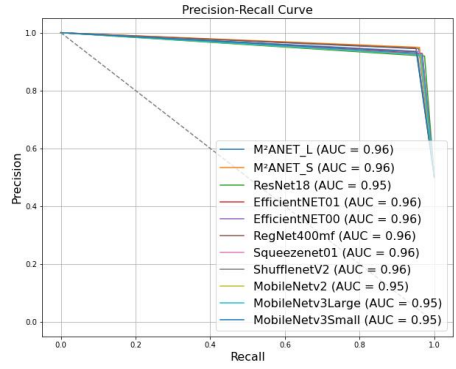

**Figure 4** ROC  **Figure 5** Precision-Recall Curve

**Computational complexity.** $M^2$ANET S and L variants, shows promising results in computational complexity compared to SOTA methods architectures. $M^2$ANET − L achieves competitive performance with relatively lower parameter count and file size compared to models like ResNet-18 and MobileNetV3-L, while maintaining efficient latency and throughput. Similarly, $M^2$ANET − S is lighter with significantly reduced latency, making it suitable for real-time applications without compromising on model accuracy. These results show the potential of $M^2$ANET as efficient alternatives for practical deployment in resource-constrained environments.

**Table 1.** Computational complexity

| Model | #Params | #FLOPs | File size | Latency | Throughput img/sec |
|---|---|---|---|---|---|
| ResNet-18 | 11.2m | 0.49 | **3.8 mb** | **0.005s** | **10401** |
| Efficient-B0 | 4.0m | **0.11** | 16.0 mb | 0.018s | 2421 |
| Efficent-B1 | 6.5m | 0.17 | 25.9 mb | 0.027s | 2357 |
| RegNet400mf | 3.9m | **0.11** | 15.5 mb | 0.021s | 1603 |
| $M^2$ANET-L | **2.2m** | 2.73 | 9.9 mb | 0.009s | 7157 |
| MobileNetV2 | 2.2m | 0.09 | 8.9 mb | 0.012s | 5551 |
| MobileNetV3-L | 4.2m | 0.06 | 16.7 mb | 0.022s | 2550 |
| MobileNetV3-S | 1.5m | **0.02** | 6.1 mb | 0.012s | 4436 |
| ShuffleNetV2 | 1.2m | 0.04 | 5.1 mb | 0.014s | 3984 |
| Squeezenet1-0 | **0.7m** | 0.17 | **2.9 mb** | 0.004s | **15557** |
| $M^2$ANET-S | 1.2m | 2.42 | 6.1 mb | **0.0015s** | 8023 |

**Classification accuracy and comparison.** Table 2 presents the top-1 accuracy and Cohen Kappa scores of various models including $M^2$ANET, showing the performance in classification tasks. Overall, $M^2$ANET-S achieved the highest top-1 accuracy of 95.45% and a Cohen Kappa score of 0.91, indicating its superior performance in accurately classifying parasitized cell images and non-parasitized images. Notably, ResNet-18, Efficient-B0, Efficient-B1, and Squeezenet1-0 also demonstrate strong performance, with top-1 accuracy scores ranging from 94.42% to 94.86% and Cohen Kappa scores around 0.89 to 0.90. These findings highlight the effectiveness of $M^2$ANET model, in achieving high accuracy and reliability in disease classification tasks of Plasmodium in cell images compared to some of the SOTA architectures such as ResNet, EfficientNet, and MobileNet variants.

**Table 2** top-1 accuracy and Cohen Kappa

| | Model | Input | epoch | F1-score | Recall | Precision | Top-1 acc. | Cohen Kappa |
|---|---|---|---|---|---|---|---|---|
| Lightweight | ResNet-18 | 112 | 90 | **0.95** | 0.97 | 0.92 | 94.42% | 0.89 |
| | Efficient-B0 | 112 | 90 | **0.95** | 0.97 | 0.93 | 94.57% | 0.89 |
| | Efficent-B1 | 112 | 90 | **0.95** | 0.97 | 0.93 | 94.64% | 0.89 |
| | RegNet400mf | 112 | 90 | **0.95** | 0.95 | **0.95** | 94.86% | **0.90** |
| | $M^2$ANET-L | 112 | 90 | **0.95** | 0.96 | 0.94 | **95.11%** | **0.90** |
| Mobile based | MobileNetV2 | 112 | 90 | 0.94 | **0.96** | 0.92 | 94.10% | 0.88 |
| | MobileNetV3-L | 112 | 90 | 0.94 | **0.96** | 0.92 | 94.27% | 0.89 |
| | MobileNetV3-S | 112 | 90 | 0.94 | 0.95 | 0.93 | 94.22% | 0.88 |
| | ShuffleNetV2 | 112 | 90 | 0.94 | 0.95 | 0.93 | 94.36% | 0.89 |
| | Squeezenet1-0 | 112 | 90 | **0.95** | 0.96 | 0.94 | 94.70% | 0.89 |
| | $M^2$ANET-S | 112 | 90 | **0.95** | 0.96 | 0.95 | **95.45%** | **0.91** |

**Sensitivity and Specificity.** Table 3 shows the performance of various models evaluated using a 5-fold cross-validation approach, with True Positive Rate (TPR) and True Negative Rate (TNR) as the key metrics. TPR is a measure of sensitivity, indicating the model's ability to correctly identify positive cases, while TNR is a measure of specificity, indicating the model's ability to correctly identify negative cases.

**Comparative analysis of models.** ResNet-18 demonstrates high sensitivity, ranging from 96.88% to 97.77%, indicating strong performance in identifying positive cases. However, its specificity ranges from 90.90% to 92.09%, which, although consistent, is lower compared to other models like $M^2$ ANET-S, indicating less accuracy in identifying negative cases.

EfficientNet-B0 shows sensitivity ranging from 96.17% to 97.34%, showing high accuracy in positive case identification, similar to ResNet-18. Specificity for EfficientNet-B0 ranges from 91.88% to 92.89%, which is slightly higher than ResNet-18, demonstrating better performance in negative case identification.

EfficientNet-B1 achieved a high consistent sensitivity, ranging from 95.95% to 97.20%, EfficientNet-B1 matches the positive case identification capabilities of the previous models. Its specificity, ranging from 91.98% to 93.26%, indicates an improvement over both ResNet-18 and EfficientNet-B0 in negative case identification.

RegNet-400MF shows lower sensitivity compared to other models, ranging from 93.83% to 95.66%. However, its specificity, ranging from 94.28% to 95.31%, is among the highest, making it excellent at identifying negative cases but less effective at identifying positive cases compared to others.

$M^2$ ANET-L achieved sensitivity from 95.32% to 96.48%, and specificity from 93.91% to 94.51%. $M^2$ ANET-L balances well between high positive case identification and good negative case identification, making it a robust model for both metrics.

The sensitivity for MobileNetV2 ranges from 95.74% to 96.84%, and specificity ranges from 90.84% to 92.82%. It performs similarly to ResNet-18 in positive case identification but has a slightly lower specificity, indicating less accuracy in identifying negative cases.

MobileNetV3-L sensitivity ranges from 95.95% to 96.79%, with specificity from 91.53% to 92.53%. While it performs well in positive case identification, it does not outperform $M^2$ANET-L or EfficientNet-B1 in negative case identification.

Sensitivity for MobileNetV3-S ranges from 94.68% to 95.74%, slightly lower than the L variant. Specificity ranges from 92.87% to 93.70%, indicating stable but not outstanding performance in negative case identification compared to larger models.

ShuffleNetV2 shows variability in sensitivity, ranging from 94.11% to 96.12%, and specificity from 92.87% to 94.14%. ShuffleNetV2 is consistent in negative case identification but shows some variation in identifying positive cases.

SqueezeNet1-0 has a sensitivity score ranging from 95.52% to 96.55%, and specificity from 92.87% to 93.85%, SqueezeNet1-0 performs similarly to MobileNetV3-L, indicating high accuracy in positive case identification and stable performance in negative case identification.

$M^2$ ANET-S stands out with sensitivity ranging from 95.75% to 96.42%, and specificity from 93.99% to 96.05%. The model not only achieves high accuracy in identifying positive cases but also shows the highest performance in negative case identification among all models, making it the most balanced and reliable model for both metrics.

Table 3 Sensitivity and Specificity using 5-fold cross validation

| | Model | k-fold 1 | | k-fold 2 | | k-fold 3 | | k-fold 4 | | k-fold 5 | |
|---|---|---|---|---|---|---|---|---|---|---|---|
| | | TPR | TNR | TPR | TNR | TPR | TNR | TPR | TNR | TPR | TNR |
| Lightweight | ResNet-18 | **96.88** | 91.46 | **97.34** | 91.58 | **97.77** | 92.09 | **97.50** | 90.90 | **97.46** | 91.25 |
| | Efficient-B0 | 96.17 | 92.58 | 96.48 | 92.01 | 97.34 | 92.89 | 97.21 | 92.33 | 96.86 | 91.88 |
| | Efficent-B1 | 95.95 | 91.98 | 96.84 | 93.26 | 97.20 | 92.67 | 97.13 | 92.54 | 96.64 | 92.24 |
| | RegNet400mf | 93.83 | **94.28** | 95.40 | **95.31** | 95.18 | **94.73** | 95.66 | 94.41 | 95.37 | **94.50** |
| | $M^2$ANET-L | 95.32 | 93.91 | 95.76 | 94.51 | 96.48 | 94.51 | 95.81 | **94.48** | 95.89 | 94.42 |
| Mobile based | MobileNetV2 | 95.81 | 91.83 | 96.33 | 90.84 | **96.84** | 92.82 | 96.32 | 92.19 | 95.74 | 92.24 |
| | MobileNetV3-L | 95.95 | 91.91 | **96.62** | 92.23 | 96.33 | 92.53 | **96.76** | 92.11 | **96.79** | 91.53 |
| | MobileNetV3-S | 94.68 | 92.87 | 95.11 | 93.48 | 95.26 | 93.70 | 95.74 | 93.12 | 95.14 | 93.08 |
| | ShufflenetV2 | 94.11 | 92.87 | 96.12 | 93.11 | 95.97 | **94.14** | 95.22 | 93.19 | 95.74 | 93.15 |
| | Squeezenet1-0 | 95.53 | 92.87 | 96.55 | 93.85 | 96.48 | 93.63 | 96.03 | 93.19 | 95.52 | 93.37 |
| | $M^2$ANET-S | 96.07 | **94.50** | 95.80 | **94.71** | 96.07 | 93.99 | 95.75 | **96.05** | 96.42 | **95.15** |

Sensitivity and specificity are key metrics for evaluating the performance of medical diagnostic models. High sensitivity is crucial for detecting as many positive cases as possible, thereby reducing the risk of missed diagnoses. High specificity ensures that negative cases are correctly identified, preventing unnecessary anxiety and treatment. A balanced approach between these metrics is essential for creating reliable and efficient diagnostic tools, particularly in resource-contained settings, like mobile and edge devices. Understanding and optimizing these metrics can significantly enhance the effectiveness of medical diagnostic systems like $M^2$ ANET in identifying conditions such as plasmodium parasitized cells.

### 4.2    Ablation

| Settings | Component | Layers | #Params | FLOPs | Accuracy |
|---|---|---|---|---|---|
| (a) | MBconv3 | [8] | 8.5m | 3.05G | 92.76 |
| (b) | MBconv3 + MHSA | [4, 4] | 2.2m | 2.73G | 95.11 |
| (d) | MBconv3 + MHSA | [4, 2] | **1.2m** | **2.42G** | **95.45** |

Using only MBConv3 layers resulted in an accuracy of 92.76%, with 8.5 million parameters and 3.05 GFLOPs. Integrating 4 MBConv3 layers with 4 MHSA layers improved accuracy to 95.11%, while reducing parameters to 2.2 million and FLOPs to 2.73 GFLOPs. Further reducing MHSA layers to 2 while maintaining 4 MBConv3 layers achieved the highest accuracy of 95.45%, with only 1.2 million parameters and 2.42 GFLOPs, indicating optimal efficiency and performance.

## 5    Conclusion

This work introduces $M^2$ANET, a novel mobile hybrid model designed to classify Plasmodium parasites in infected cell images. The design skillfully integrates MBConv3 and a modified 2D MHSA, optimized for compatibility with resource constrained setting like mobile and edge devices. Comprehensive evaluation and comparison with some state-of-the-art models demonstrate its effectiveness in identifying infected cell images. The model has the potential to improve malaria diagnosis, especially in resource constrained settings. Further research should explore its applicability in real-world clinical settings and assess its scalability for large-scale deployment.

**Disclosure of Interests.** The authors have no competing interests to declare that are relevant to the content of this article.

**Funding.** This research received no external funding.